\documentclass[11pt,a4paper,english,showpacs]{revtex4}

\usepackage{amssymb}
\usepackage{amsmath}
\usepackage{graphicx}
\usepackage{babel}
\usepackage{hyperref}

\begin{document}

\title{Spontaneous Gauge Symmetry Breaking in a SUSY Chern-Simons Model}

\author{A.~C.~Lehum}
\author{A.~F.~Ferrari}
\author{M.~Gomes}
\author{A.~J.~da~Silva}
\affiliation{Instituto de F\'{\i}sica, Universidade de S\~{a}o Paulo\\
 Caixa Postal 66318, 05315-970, S\~{a}o Paulo - SP, Brazil}
\email{lehum, alysson, mgomes, ajsilva@fma.if.usp.br}


\begin{abstract}
This works presents a perturbative analysis of the supersymmetric Chern-Simons model in three spacetime dimensions coupled to a Higgs field, using the superfield formalism. We study the spontaneous symmetry breaking of the $U(1)$ gauge symmetry and evaluate the first quantum corrections to the effective action in the broken phase. We show that the infinite renormalization of the gap equation is enough to ensure the renormalizability of the model at the first loop level. 
\end{abstract}

\pacs{1.15.Ex, 11.10.Gh, 11.15.-q,11.30.Pb,12.60.Jv}

\keywords{Chern-Simons, spontaneous symmetry breaking, supersymmetry, superfields}

\maketitle

Field theories in three-dimensional spacetime are often simpler than their similar four dimensional counterparts and, as such, can be regarded as useful laboratories for several field theoretical properties. In particular, induced by the Chern-Simons term~\cite{Deser:1981wh,Schonfeld:1981}, three dimensional theories exhibit massive gauge fields, exotic statistics and fractional spin, relevant qualities for the study of the quantized Hall effect~\cite{prange}. In the non-Abelian case, the invariance of the action under large gauge transformations requires the Chern-Simons coefficient to be quantized~\cite{Deser:1981wh}, an aspect that can be explicitly verified in perturbation theory~\cite{Pisarski:1985yj,Giavarini:1992xz}.

One interesting possibility that has been considered in the literature is the coupling of the Chern-Simons term to a Higgs field (CSH), thus allowing for spontaneous gauge symmetry breaking to occur, and a nontrivial dynamics  for the Chern-Simons gauge field to be settled (giving rise to a self-dual model, which happens to be equivalent to a Maxwell-Chern-Simons theory~\cite{Deser:1989gf}). The quantization of the Chern-Simons coefficient for non-Abelian theories in the broken phase holds if the remaining gauge symmetry is non-Abelian~\cite{Chen:1994zx,Khare:1993ma}, whereas such quantization does not happen in the case of an Abelian theory, or a completely broken non-Abelian gauge symmetry~\cite{Khlebnikov:1990ai,Spiridonov:1990st,Khlebnikov:1991gg}. Also, for a specific form of the Higgs potential, the CSH model has solutions which satisfy a Bogomol'nyi equation~\cite{Jackiw:1990aw,Hong:1990yh}. The exact form of this potential can be obtained either by imposing a self-dual condition on the matter field~\cite{Srivastava:1990cw} or by enlarging the model to obtain an ${\cal N} = 2$ supersymmetric theory~\cite{Lee:1990it}. It is interesting to note that a tridimensional analog of the Coleman-Weinberg model in four spacetime dimensions~\cite{Coleman:1973jx} is realized by a Chern-Simons field minimally coupled to a massless scalar field with a purely sextuple self-interaction. In this case, however, dynamical symmetry breaking appears only at two loops~\cite{Tan:1996kz,Tan:1997ew,deAlbuquerque:2000ec,Dias:2003pw}, and not at the one loop level, as in four dimensions. 

In this work, we will investigate some perturbative properties of the ${\cal N} = 1$ supersymmetric Chern-Simons-Higgs model in $2+1$ dimensions. We use the superfield formalism, so that supersymmetry is manifestly preserved, and study the phase structure and the renormalizability of the model at the one-loop level. 

Our starting point is the action of the supersymmetric Chern-Simons model coupled to a scalar superfield $\Phi$,
\begin{eqnarray}\label{eq1}
S=\int{d^5z}\Big{\{}A^{\alpha}W_{\alpha}
-{1\over 2}\overline{\nabla^{\alpha}\Phi}\nabla_{\alpha}\Phi
+\mu\bar\Phi\Phi-\lambda(\bar\Phi\Phi)^2\Big{\}}~,
\end{eqnarray}

\noindent
where  $A_\alpha$ is the gauge superpotential, $W^{\alpha}=(1/2)D^{\beta}D^{\alpha}A_{\beta}$ is the covariant field strength, $\nabla^{\alpha}=(D^{\alpha}-ieA^{\alpha})$ is the gauge supercovariant derivative, and $D_{\alpha} = \partial_{\alpha} + i \theta^{\beta} \partial_{\alpha \beta}$ is the usual supersymmetric covariant derivative (in this paper, we follow the conventions of~\cite{Gates:1983nr}). This action is invariant under the infinitesimal gauge transformations,
\begin{eqnarray}\label{eq2}
\bar\Phi\longrightarrow \bar\Phi^{\prime}=\bar\Phi(1-ieK)~,\nonumber\\
\Phi\longrightarrow \Phi^{\prime}=(1+ieK)\Phi~,\\
A_{\alpha}\longrightarrow A_{\alpha}^{\prime}=A_{\alpha}+D_{\alpha}K~,\nonumber
\end{eqnarray}

\noindent
where $K=K(x,\theta)$ is a real scalar superfield playing the role of the gauge parameter.

For positive values of the parameter $ \mu $, the classical potential
\begin{eqnarray}\label{eq3}
V(\bar\Phi,\Phi)=-\mu\bar\Phi\Phi+\lambda(\bar\Phi\Phi)^2
\end{eqnarray}

\noindent
has a non-trivial minimum specified by
\begin{eqnarray}\label{eq4}
|\Phi|=\sqrt{\mu\over 2\lambda}~.
\end{eqnarray}

\noindent
With this in mind, we perform a shift $v$ in the superfield $ \Phi $ ($ \bar\Phi $),
\begin{eqnarray}\label{eq5}
\Phi&=&{1\over\sqrt{2}}\Big(\Sigma+v+i\Pi\Big)~,\nonumber\\
\bar\Phi&=&{1\over\sqrt{2}}\Big(\Sigma+v-i\Pi\Big)\,,
\end{eqnarray}

\noindent
in order to write an action in terms of the real superfields $\Sigma$ and $\Pi$ which satisfy $<\Sigma>=<\Pi>=0$. The gauge transformation for $\Sigma$ and $\Pi$ are given by
\begin{eqnarray}\label{eq5a}
\Pi &\longrightarrow &\Pi^{\prime}=\Pi+eK(\Sigma+v)~,\nonumber\\
\Sigma & \longrightarrow & \Sigma^{\prime}=\Sigma-eK\Pi~.
\end{eqnarray}

\noindent
Thus, the action in terms of the real superfields is given by
\begin{eqnarray}\label{eq6}
S_1&=&\int{d^5z}\Big{\{}A^{\alpha}W_{\alpha}-{e^2v^2\over 4}A^{\alpha}A_{\alpha}
+{ev\over 2}D^{\alpha}A_{\alpha}\Pi
+{1\over 2}\Sigma[D^2-(3\lambda v^2-\mu)]\Sigma\nonumber\\
&+&{1\over 2}\Pi[D^2-(\lambda v^2-\mu)]\Pi+{e\over 2}D^{\alpha}\Pi A_{\alpha}\Sigma
-{e\over 2}D^{\alpha}\Sigma A_{\alpha}\Pi
-{e^2\over 2}(\Sigma^2+\Pi^2)A^2\\
&-&ve^2\Sigma A^2-{\lambda\over 4}(\Sigma^4+\Pi^4)
-{\lambda\over 2}\Sigma^2\Pi^2-\lambda v\Sigma(\Sigma^2+\Pi^2)
+v(\mu-v^2\lambda)\Sigma\Big{\}}~.\nonumber
\end{eqnarray}

To eliminate the mixing between $A^{\alpha}$ and $\Pi$ that appears in Eq.~(\ref {eq6}), we use an $R_{\xi}$ gauge fixing depending on a gauge parameter $\alpha$, which, together with the corresponding Faddeev-Popov determinant, is introduced through the action
\begin{eqnarray}\label{eq6a}
S_{GF+FP}&=&\int{d^5z}\Big[-{1\over 2\alpha}
(D^{\alpha}A_{\alpha}+\alpha{ev\over 2}\Pi)^2
-\bar{c}D^2c+{\alpha\over 4}e^2v^2\bar{c}c+{\alpha\over 4}{e^2v}\bar{c}\Sigma c\Big]~,
\end{eqnarray}

\noindent
where $c, \bar{c}$ are scalar ghost superfields. By adding Eqs.~(\ref{eq6}) and~(\ref{eq6a}), we obtain 
\begin{eqnarray}\label{eq6b}
S_2&=&\int{d^5z}\Big{\{}A^{\alpha}W_{\alpha}-M_{A}A^{\alpha}A_{\alpha}
-{1\over 2\alpha}D^{\alpha}A_{\alpha}D^{\beta}A_{\beta}
+{1\over 2}\Sigma(D^2-M_{\Sigma})\Sigma\nonumber\\
&+&{1\over 2}\Pi(D^2-M_{\Pi})\Pi+{e\over 2}D^{\alpha}\Pi A_{\alpha}\Sigma
-{e\over 2}D^{\alpha}\Sigma A_{\alpha}\Pi
-{e^2\over 2}(\Sigma^2+\Pi^2)A^2-ve^2\Sigma A^2\\
&-&{\lambda\over 4}(\Sigma^4+\Pi^4)+v(\mu-v^2\lambda)\Sigma
-{\lambda\over 2}\Sigma^2\Pi^2-\lambda v\Sigma(\Sigma^2+\Pi^2)
-\bar{c}(D^2-M_c)c+{\alpha\over 4}e^2v\bar{c}\Sigma{c}
\Big{\}}~.\nonumber
\end{eqnarray}

\noindent
where $M_{\Sigma}=(3\lambda v^2-\mu)$, $M_{A}=e^2v^2/4$, $M_{\Pi}=(\lambda v^2-\mu+\alpha M_{A})$ and 
$M_c=\alpha M_{A}$.

In the minimum of the classical potential, $v^2=\mu / \lambda$, and the mechanism of spontaneous breakdown of symmetry generates masses $M_{\Sigma}=2\mu$ and $ M_{A}=e^2 \mu/4\lambda$ for the superfields $\Sigma$ and $A^{\alpha}$, respectively. The superfield $\Pi$ also acquires mass, but only from the process of gauge fixing, since its mass turns out to depend on the gauge parameter $\alpha$. Therefore, $\Pi$ must be an unphysical field, which should not be observed in external legs of any scattering process~\cite{Peskin:1995ev}.

From Eq.~(\ref{eq6b}), one readily obtains the propagators of the model,
\begin{eqnarray}\label{eq7}
\langle T\,\Sigma(k,\theta_1)\Sigma(-k,\theta_2)\rangle&=&-i{D^2+M_{\Sigma}\over k^2+M_{\Sigma}^2}\delta^{(2)}(\theta_1-\theta_2)~,\nonumber\\
\langle T\,\Pi(k,\theta_1)\Pi(-k,\theta_2)\rangle&=&-i{D^2+M_{\Pi}\over k^2+M_{\Pi}^2}\delta^{(2)}(\theta_1-\theta_2)~,\nonumber\\
\langle T\,c(k,\theta_1)\bar{c}(-k,\theta_2)\rangle&=&+i{(D^2+M_c)\over k^2+M_c^2}\delta^{(2)}(\theta_1-\theta_2)~,\\
\langle T\,A_{\alpha}(k,\theta_1)A_{\beta}(-k,\theta_2)\rangle&=&{i\over 4}\Big[{(D^2+M_{A})D^2D_{\beta}D_{\alpha}\over k^2(k^2+M_{A}^2)}\nonumber\\
&+&\alpha{(D^2-\alpha M_{A})D^2D_{\alpha}D_{\beta}\over k^2(k^2+\alpha^2M_{A}^2)}\Big]\delta^{(2)}(\theta_1-\theta_2)~.\nonumber
\end{eqnarray}

\noindent
In the process of renormalization, we need to redefine the superfields as
$\bar\Phi\rightarrow Z_1^{1/2}\bar\Phi$, $\Phi\rightarrow Z_1^{1/2}\Phi$, $c\rightarrow Z_c^{1/2}c$ and 
$A^{\alpha}\rightarrow Z_2A^{\alpha}$, and also renormalize the couplings according to 
$\mu\rightarrow(\mu+\delta_{\mu})$, $e\rightarrow(e+\delta_{e})$ and 
$\lambda\rightarrow(\lambda+\delta_{\lambda})$. After this, we can write explicitly the action of counterterms for this model, 
\begin{eqnarray}\label{eq9}
S_{CT}&=&\int{d^5z}\Big{\{}{(2\delta_2+\delta_2^2)\over 2}A^{\alpha}W_{\alpha}
-{v^2\over 2}[(1+\delta_2)^2(1+\delta_1)(e+\delta_e)^2-e^2]A^2\nonumber\\
&-&{1\over 2\alpha}[(1+\delta_{2})^2-1]D^{\alpha}A_{\alpha}D^{\beta}A_{\beta}
-\delta_c~\bar{c}D^2c+{\alpha v^2\over 4}[(1+\delta_c)(e+\delta_e)^2(1+\delta_1)-e^2]\bar{c}c\nonumber\\
&+&{v\over 2}[(e+\delta_e)(1+\delta_1)(1+\delta_2)-e]D^{\alpha}A_{\alpha}\Pi
+{\delta_1\over 2}(\Sigma D^2\Sigma+\Pi D^2\Pi)\nonumber\\
&-&{1\over 2}[3(\lambda+\delta_{\lambda})(1+\delta_1)^2v^2
-(\mu+\delta_{\mu})(1+\delta_1)-M_{\Sigma}]\Sigma^2\nonumber\\
&-&{1\over 2}[(\lambda+\delta_{\lambda})(1+\delta_1)^2v^2
-(\mu+\delta_{\mu})(1+\delta_1)+{\alpha v^2\over 2}(e+\delta_e)^2(1+\delta_1)^2-M_{\Pi}]\Pi^2\nonumber\\
&+&{1\over 2}[(e+\delta_e)(1+\delta_1)(1+\delta_2)-e]
(D^{\alpha}\Pi A_{\alpha}\Sigma-D^{\alpha}\Sigma A_{\alpha}\Pi)\\
&-&{1\over 2}[(e+\delta_e)^2(1+\delta_1)(1+\delta_2)^2-e](\Sigma^2+\Pi^2)A^2\nonumber\\
&+&{\alpha\over 4}[(e+\delta_e)^2(1+\delta_1)(1+\delta_c)-e^2]\bar{c}\Sigma c\nonumber\\
&-&v[(e+\delta_e)^2(1+\delta_1)(1+\delta_2)^2-e]\Sigma A^2
-{1\over 4}[(\lambda+\delta_{\lambda})(1+\delta_1)^2-\lambda](\Sigma^4+\Pi^4)\nonumber\\
&-&{1\over 2}[(\lambda+\delta_{\lambda})(1+\delta_1)^2-\lambda]\Sigma^2\Pi^2
-v[(\lambda+\delta_{\lambda})(1+\delta_1)^2-\lambda]\Sigma(\Sigma^2+\Pi^2)\nonumber\\
&+&v[(1+\delta_1)(\mu+\delta_{\mu})
-(1+\delta_1)^2(\lambda+\delta_{\lambda})v^2-(\mu-v^2\lambda)]\Sigma\Big{\}}~,\nonumber
\end{eqnarray}

\noindent
where $\delta_1=(Z_1-1)$, $\delta_2=(Z_2-1)$ and $\delta_c=(Z_c-1)$. 

We shall begin our analysis by looking at the phase structure of the model. To this end, we investigate the one point function, whose vanishing yields the gap equation. At tree level, the gap equation is given by
\begin{eqnarray}\label{eq9b}
v(\mu-v^2\lambda)=0~,
\end{eqnarray}

\noindent
which admits two solutions, $v=0$ and $v^2=\mu/\lambda$. The first is the trivial solution where the $U(1)$ gauge symmetry is not broken. For the second solution, the symmetry $U(1)$ is broken, yet for both of these solutions, supersymmetry is explicitly maintained. We focus our attention on the renormalizability of the phase where gauge symmetry is broken, in which the Chern-Simons field acquires a non-trivial dynamics, so we will set $v^2=\mu/\lambda$ from now on.

\begin{figure}[ht]
\includegraphics[]{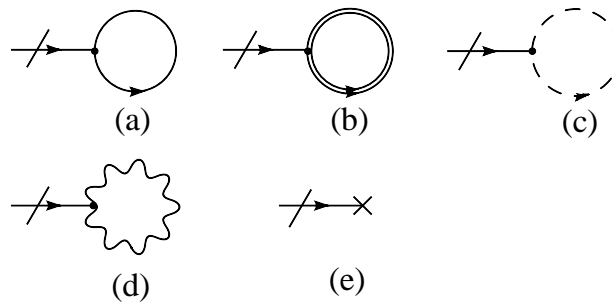}
\caption{One loop contribution to the gap equation. Continuous lines represents the $\Sigma$ propagator, double continuous lines the $\Pi$ propagator, dashed lines the ghost propagator and wavy lines the gauge superpotential propagator. The cross represents the insertion of counterterms.}  \label{Fig1}
\end{figure}

Let us first look at the quantum corrections to the gap equation. Up to one loop order, five graphs contribute to it, those depicted in Fig.~\ref{Fig1}. The resulting gap equation reads
\begin{eqnarray}\label{eq9a}
&-&3\lambda v\int{d^3k\over(2\pi)^3}{1\over k^2+M_{\Sigma}^2}
-\lambda v\int{d^3k\over(2\pi)^3}{1\over k^2+M_{\Pi}^2}\nonumber\\
&-&{\alpha\over 4}e^2v\int{d^3k\over(2\pi)^3}{1\over k^2+M_{c}^2}
-{e^2v\over 4}\int{d^3k\over(2\pi)^3}\Big[{1\over k^2+M_{A}^2}
-{\alpha\over k^2+\alpha^2M_{A}^2}\Big]\\
&&+iv[(1+\delta_1)(\mu+\delta_{\mu})
-(1+\delta_1)^2(\lambda+\delta_{\lambda})v^2-(\mu-v^2\lambda)]=0\,.\nonumber
\end{eqnarray}
 
\noindent
We calculate the divergent integrals in Eq.~(\ref{eq9a}) with the help of an ultraviolet (UV) regulator $\Lambda$, obtaining the following divergent part (note the cancellation of the $\alpha$-dependent divergent parts)
\begin{eqnarray}\label{eq10}
&&-iv{\Lambda\over 2\pi^2}\Big[4\lambda+{e^2\over 4}\Big]+iv\delta_3=0~,
\end{eqnarray}

\noindent
where 
\begin{equation}
\label{delta3}
\delta_3=[(1+\delta_1)(\mu+\delta_{\mu})
-(1+\delta_1)^2(\lambda+\delta_{\lambda})v^2-(\mu-v^2\lambda)]=\delta_\mu - \mu \delta_1 - v^2 \delta_\lambda \,.
\end{equation}

\begin{figure}[ht]
\includegraphics[]{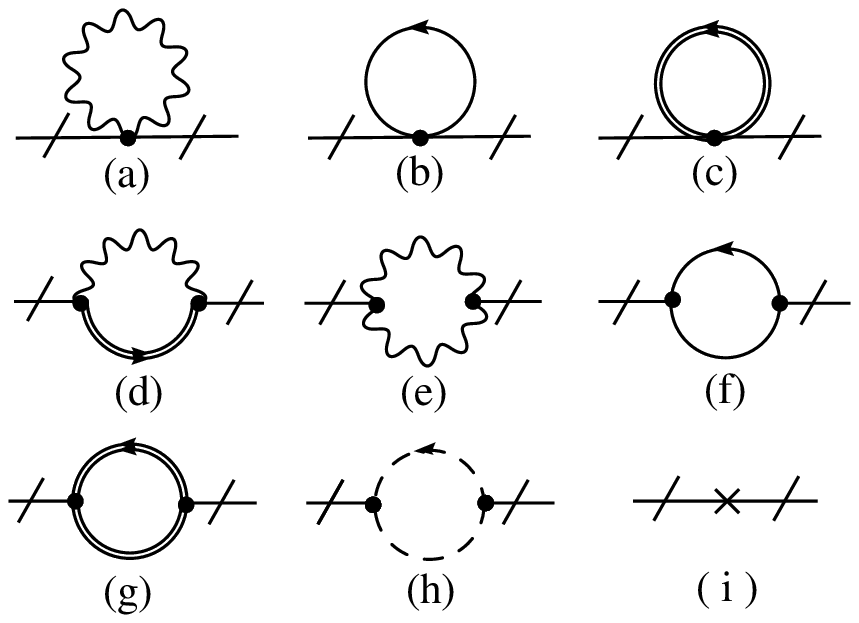}
\caption{One loop contribution to the self-energy of $\Sigma$ superfield.}   \label{Fig2}
\end{figure}

Now we turn to the investigation of the divergence properties of the two-point vertex functions. The diagrams contributing to the two-point function of the $\Sigma$ superfield are drawn in Fig.~\ref{Fig2}. We shall be mainly interested in the evaluation of divergent contributions, which can come from the the first four graphs in Fig.~\ref{Fig2}. The contribution to $S_{\Sigma\Sigma}$ from the graph \ref{Fig2}a is given by
\begin{eqnarray}\label{eq12}
S_{\Sigma\Sigma~a}&=&-{e^2\over 8}\int\!{d^3p\over(2\pi)^3}d^2\theta\,\Sigma(p,\theta)\Sigma(-p,\theta)C^{\alpha\beta}
\int\!{d^3k\over (2\pi)^3}\nonumber\\
&\times&\Big[ \frac{D^2(D^2+M_{A})D_{\beta}D_{\alpha} \delta_{\theta\theta} }{k^2(k^2+M_{A}^2)}
+ \alpha \frac{D^2(D^2-\alpha M_{A})D_{\alpha}D_{\beta} \delta_{\theta\theta} }{k^2(k^2+\alpha^2M_{A}^2)}\Big]~,
\end{eqnarray}

\noindent
where $\delta_{\theta\theta}$ means the limit of $\delta^{(2)}(\theta-\theta^\prime)$ when $\theta^\prime \rightarrow \theta$ after the application of the covariant derivatives. After standard D-algebra manipulations~\cite{footnote}, expression~(\ref{eq12}) results in
\begin{eqnarray}\label{eq12b}
S_{\Sigma\Sigma~a}&=&-{e^2\over 4}\int\!{d^3p\over(2\pi)^3}d^2\theta \,
\Sigma(p,\theta)\,\Sigma(-p,\theta)\int\!{d^3k\over (2\pi)^3}\,\left(
{1\over k^2+M_{A}^2}-{\alpha\over k^2+\alpha^2M_{A}^2}\right)~.
\end{eqnarray}

\noindent
For the contributions from graphs \ref{Fig2}b and \ref{Fig2}c, one obtains
\begin{eqnarray}\label{eq13}
S_{\Sigma\Sigma~bc}&=&-\int\!{d^3p\over(2\pi)^3} \, d^2 \theta \, \Sigma(p,\theta)\Sigma(-p,\theta) 
\int\!{d^3k\over (2\pi)^3}
\Big{\{}{3\lambda}{(D^2+M_{\Sigma})\delta_{\theta\theta}\over k^2+M_{\Sigma}^2}
+{\lambda}{(D^2+M_{\Pi})\delta_{\theta\theta}\over k^2+M_{\Pi}^2}\Big{\}}~.
\end{eqnarray}

\noindent
As for diagram \ref{Fig2}d, we have
\begin{eqnarray}\label{eq13a}
S_{\Sigma\Sigma~d}=-{e^2\over 8}\int\!{d^3p\over(2\pi)^3}d^2\theta_1d^2\theta_2
\int\!{d^3k\over (2\pi)^3}{C_{\alpha\beta}\over[(k+p)^2+M_{\Pi}^2]}
~~\hspace{5cm}~\\
\times\Big{\{} D_1^{\alpha}(D^2+M_{\Pi})\delta_{12}
\Big[{D^2(D^2+M_{A})D_{1\beta}D_{1\alpha}\over k^2(k^2+M_{A}^2)}
+\alpha{D^2(D^2-\alpha M_{A})D_{1\alpha}D_{1\beta}\over k^2(k^2+\alpha^2M_{A}^2)}\Big]\delta_{12}
\Sigma(p,\theta_1)D_2^{\beta}\Sigma(-p,\theta_2)\nonumber\\
+D_2^{\beta}(D^2+M_{\Pi})\delta_{12}
\Big[{D^2(D^2+M_{A})D_{1\beta}D_{1\alpha}\over k^2(k^2+M_{A}^2)}
+\alpha{D^2(D^2-\alpha M_{A})D_{1\alpha}D_{1\beta}\over k^2(k^2+\alpha^2M_{A}^2)}\Big]
\delta_{12}D_1^{\alpha}\Sigma(p,\theta_1)\Sigma(-p,\theta_2)\nonumber\\
-D_1^{\alpha}D_2^{\beta}(D^2+M_{\Pi})\delta_{12}
\Big[{D^2(D^2+M_{A})D_{1\beta}D_{1\alpha}\over k^2(k^2+M_{A}^2)}
+\alpha{D^2(D^2-\alpha M_{A})D_{1\alpha}D_{1\beta}\over k^2(k^2+\alpha^2M_{A}^2)}\Big]
\delta_{12}\Sigma(p,\theta_1)\Sigma(-p,\theta_2)\nonumber\\
-(D^2+M_{\Pi})\delta_{12}
\Big[{D^2(D^2+M_{A})D_{1\beta}D_{1\alpha}\over k^2(k^2+M_{A}^2)}
+\alpha{D^2(D^2-\alpha M_{A})D_{1\alpha}D_{1\beta}\over k^2(k^2+\alpha^2M_{A}^2)}\Big]
\delta_{12}D_1^{\alpha}\Sigma(p,\theta_1)D_2^{\beta}\Sigma(-p,\theta_2)\Big{\}}~,\nonumber
\end{eqnarray}

\noindent
where the shorthand notations $\delta_{12}=\delta^{(2)}(\theta_1-\theta_2)$ and $D_i^{\alpha}=D^{\alpha}(\theta_i)$ were used. The complete evaluation of Eq.~(\ref{eq13a}) yields,
\begin{eqnarray}\label{eq13b}
S_{\Sigma\Sigma d}&=&-\alpha{e^2\over 4}
\int\!{d^3p\over(2\pi)^3}d^2 \theta \,
\Sigma(p,\theta)\Sigma(-p,\theta)\nonumber\\
&\times&\Big{\{}\int\!{d^3k\over (2\pi)^3}{k^2\over[(k+p)^2+M_{\Pi}^2](k^2+\alpha^2M_A^2)}+(\textrm{finite terms})\Big{\}}~.
\end{eqnarray}

The momentum integral in Eqs.~(\ref{eq12b}), (\ref{eq13}) and (\ref{eq13b}) is again performed with an UV regulator $\Lambda$, resulting in the following correction to the effective action of $\Sigma^2$,
\begin{equation}\label{eq14}
S_{\Sigma\Sigma}=-\int{d^3p\over(2\pi)^3}d^2\theta \, \Sigma(p,\theta) \, \Big{\{}i\Big[4\lambda+{e^2\over 4}\Big]{\Lambda\over 2\pi^2}+i\delta_4-i\frac{\delta_1}{2}D^2+(\textrm{finite terms})\Big{\}}\Sigma(-p,\theta)~,
\end{equation}

\noindent
where $\delta_{4}=
[3(\lambda+\delta_{\lambda})(1+\delta_1)^2v^2-(\mu+\delta_{\mu})(1+\delta_1)-M_{\Sigma}]=2v^2\delta_{\lambda}-\delta_3$. We verify that the divergence in $S_{\Sigma\Sigma}$ is completely eliminated by the $\delta_3$ counterterm, already fixed by the gap equation~(\ref{eq10}). The conclusion is that the counterterm $\delta_{\lambda}$ must be finite. From the absence of wave function renormalization, the counterterm $\delta_1$ is also finite, so that from Eq.~(\ref{delta3}), only $\delta_\mu$ needs to be divergent.

\begin{figure}[ht]
\includegraphics[]{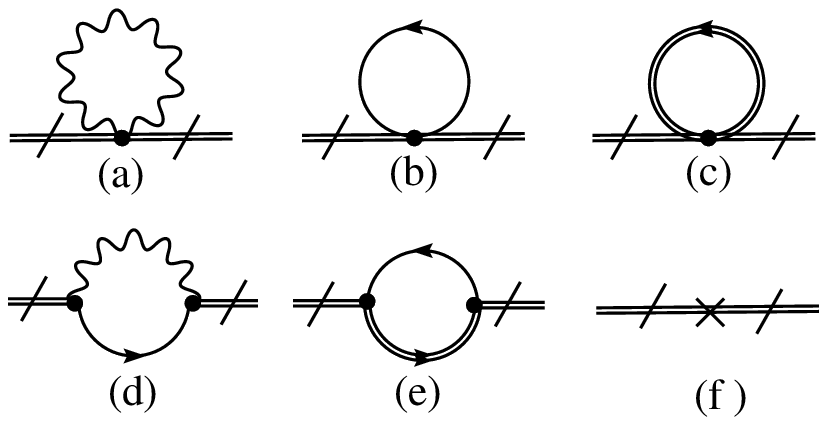}
\caption{One loop contribution to the self-energy of $\Pi$ superfield.} \label{Fig3}
\end{figure}

The diagrams that contribute to the radiative corrections to the self-energy of the $\Pi$ superfield are represented in Fig.~\ref{Fig3}, and again only the first four graphs are ultraviolet divergent. Interestingly enough, the graphs~\ref{Fig3}a to~\ref{Fig3}d are identical to the corresponding ones in Fig.~\ref{Fig2}, after interchanging $\Pi\leftrightarrow\Sigma$. Because of this, we can use our result in Eq.~(\ref{eq14}) to write directly
\begin{equation}
\label{eq15}
S_{\Pi\Pi}=-\int{d^3p\over(2\pi)^3}d^2\theta \, \Pi(p,\theta) \, \Big{\{}i\Big[4\lambda+{e^2\over 4}\Big]{\Lambda\over 2\pi^2}+i\delta_{5}-i\frac{\delta_1}{2}D^2+(\textrm{finite terms})\Big{\}}\Pi(-p,\theta)~,
\end{equation}

\noindent
where 
\begin{eqnarray}
\delta_5&=&[(\lambda+\delta_{\lambda})(1+\delta_1)^2v^2
-(\mu+\delta_{\mu})(1+\delta_1)+{\alpha v^2\over 2}(e+\delta_e)^2(1+\delta_1)^2-M_{\Pi}]\nonumber\\
&=&-\delta_3+\frac{\alpha v^2}{2}(\delta_e+e^2\delta_1).
\end{eqnarray} 

\noindent
Again, all ultraviolet divergences in this correction are cancelled by the gap equation renormalization, so that the counterterm $\delta_e$ is finite.

\begin{figure}[ht]
\includegraphics[]{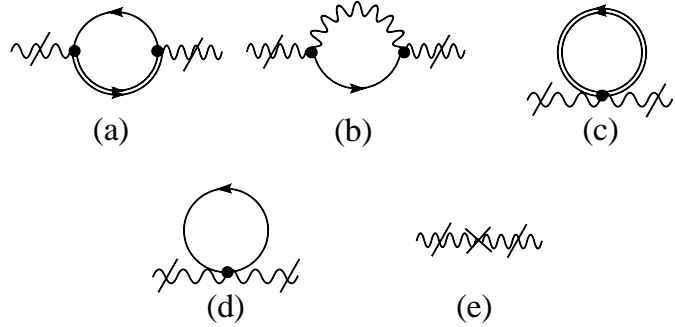}
\caption{One loop contribution to the self-energy of gauge superfield $A_{\alpha}$.}    \label{Fig4}
\end{figure}

As for the effective action of the gauge superfield, four diagrams contribute to its quadratic part, which are those in Fig.~\ref{Fig4}. The contribution of diagram \ref{Fig4}a is given by
\begin{eqnarray}\label{eq17b}
S_{AAa}&=& {e^2\over 8}\int\!{d^3p\over(2\pi)^3}d^2\theta~\int\!{d^3k\over(2\pi)^3}
{1\over[(k+p)^2+M_{\Pi}^2](k^2+M_{\Sigma}^2)}\nonumber\\
&\times&A^{\alpha}(p,\theta)\Big{\{}3C_{\beta\alpha}p^2-3Mp_{\alpha\beta}+(M_{\Sigma}+M_{\Pi})C_{\beta\alpha}D^2\\
&+&4M_{\Sigma}M_{\Pi}C_{\beta\alpha}+p_{\beta\alpha}D^2
+5Np_{\alpha\beta}+4k^2C_{\beta\alpha}\Big{\}}A^{\beta}(-p,\theta).\nonumber
\end{eqnarray}

\noindent
After some algebraic manipulation, this expression can be cast as
\begin{eqnarray}\label{eq17}
S_{AAa}&=& -{e^2\over 8}\int\!{d^3p\over(2\pi)^3}d^2\theta~\int\!{d^3k\over(2\pi)^3}
{1\over[(k+p)^2+M_{\Pi}^2](k^2+M_{\Sigma}^2)}\nonumber\\
&\times&\Big{\{}4W^{\alpha}W_{\alpha}-4{(M_{\Sigma}-M_{\Pi})}A^{\alpha}W_{\alpha}
+4M_{\Sigma}M_{\Pi}A^{\alpha}A_{\alpha}\\
&-&D^{\alpha}A_{\alpha}\big[D^2-{2(M_{\Sigma}-M_{\Pi})}\big]D^{\beta}A_{\beta}
-4k^2A^{\alpha}A_{\alpha}\Big{\}}.\nonumber
\end{eqnarray}

\noindent
We witness here a general property of the Chern-Simons field coupled to matter, that is the generation, at the quantum level, of a non-local Maxwell term $W^{\alpha}W_{\alpha}$ in the effective action of gauge superfield, which was not present in original action. As we will show below, the linear divergence in the last term in Eq.~(\ref{eq17}) will be cancelled when summing up the other graphs in Fig.~\ref{Fig4}.

The contributions of diagram~\ref{Fig4}b,
\begin{eqnarray}\label{eq17a}
S_{AAb}&=&-{v^2e^4\over 4}\int{d^3p\over(2\pi)^3}d^2\theta\int{d^3k\over(2\pi)^3}
{1\over k^2(k^2+M_{A}^2)(k^2+\alpha^2M_{A}^2)[(k+p)^2+M_{\Sigma}^2]}\nonumber\\
&\times&\Big{\{}-\alpha(k^2+M_{A}^2)\Big[-4M_{A}W^{\alpha}W_{\alpha}
+2\alpha M_{\Sigma}M_{A}A^{\alpha}W_{\alpha}\nonumber\\
&+&\alpha M_{A}D^{\alpha}A_{\alpha}(D^2-M_{\Sigma})D^{\beta}A_{\beta}
-2k^2(M_{\Sigma}-\alpha M_A)A^{\alpha}A_{\alpha}
-2k^2D^{\alpha}A_{\alpha}D^{\beta}A_{\beta}\Big]\\
&+&(k^2+\alpha^2M_{A}^2)\Big[4M_{A}W^{\alpha}W_{\alpha}
-2(k^2-M_{\Sigma}M_{A})A^{\alpha}W_{\alpha}\nonumber\\
&+&M_{A}D^{\alpha}A_{\alpha}(D^2-M_{\Sigma})D^{\beta}A_{\beta}
-2k^2(M_{\Sigma}-\alpha M_A)A^{\alpha}A_{\alpha}\Big]\Big{\}}.\nonumber
\end{eqnarray}

\noindent
and of the graphs \ref{Fig4}c and \ref{Fig4}d,
\begin{eqnarray}\label{eq18}
S_{AAcd}=-{e^2\over 4}\int\!{d^3p\over(2\pi)^3}d^2\theta~A^{\alpha}(p,\theta)A_{\alpha}(-p,\theta)\int\!{d^3k\over(2\pi)^3}\Big{\{}{1\over k^2+M_{\Sigma}^2}+{1\over k^2+M_{\Pi}^2}\Big{\}}\,,
\end{eqnarray} 

\noindent
are added to Eq.~(\ref{eq17b}), together with the contribution of the counterterms, to obtain,
\begin{eqnarray}\label{eq19}
S_{AA}&=&{e^2\over 2}\int{d^3p\over(2\pi)^3}\,d^2\theta \, 
A^{\alpha}(p,\theta)\Big{\{}
-iv^2 {\delta_{\alpha}}^{\beta} \, \delta_6
+i\delta_2\frac{D^{\beta}D_{\alpha}}{e^2}\nonumber\\
&+&{\delta_{\alpha}}^{\beta}\int{d^3k\over(2\pi)^3}{k^2\over
 (k^2+M_{\Sigma}^2)[(k+p)^2+M_{\Pi}^2]}\\
&-&{{\delta_{\alpha}}^{\beta}\over
 2}\int{d^3k\over(2\pi)^3}\Big({1\over k^2+M_{\Pi}^2}
+{1\over k^2+M_{\Sigma}^2}\Big)
+{({\rm finite})_{\alpha}}^{\beta}
\Big{\}}A_{\beta}(-p,\theta)
\nonumber
\end{eqnarray}

\noindent
where $\delta_6=[(1+\delta_2)^2(1+\delta_1)(e+\delta_e)^2-e^2]=e^2(\delta_1+2\delta_2)+2\delta_e$. We extract from Eq.~(\ref{eq19}) the divergent parts using a power expansion of the integrands around $p=0$, and observe the total cancellation of the divergences in the two point function of the gauge superfield. As a consequence, the $\delta_2$ counterterm must be finite.

\begin{figure}[ht]
\includegraphics[]{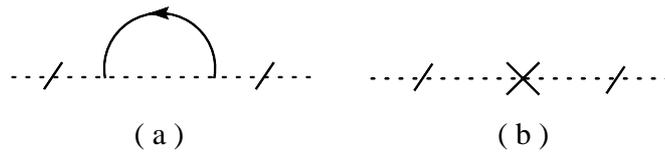}
\caption{Contribution at one loop to the ghost effective action.}  \label{Fig5}
\end{figure}

\begin{figure}[ht]
\includegraphics[]{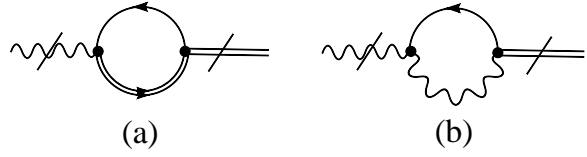}
\caption{Diagrams that mix the $A^{\alpha}$ and  $\Pi$ superfields.}  \label{Fig6}
\end{figure}

Two more remarks remain to be made concerning two point functions. First, the quadratic part of the ghost effective action also gets a quantum correction, given by the graphs in Fig.~\ref{Fig5}, 
\begin{eqnarray}\label{eq20g}
S_{\bar{c}c}&=&-{\alpha^2\over 8}e^4v^2\int{{d^3p}\over(2\pi)^3}\,d^2\theta~
\int{{d^3k}\over(2\pi)^3}{\bar{c}(-p,\theta)[D^2+(M_c+M_{\Sigma})]c(p,\theta)\over(k^2+M_c^2)[(k-p)^2+M_{\Sigma}^2]}\nonumber\\
&-&\int{{d^3p}\over(2\pi)^3}d^2\theta~\bar{c}(-p,\theta)\left[\delta_cD^2
-\frac{\alpha v^2}{4}(e^2\delta_c+e^2\delta_1+2e^2\delta_2+2\delta_e)\right]c(p,\theta)~,
\end{eqnarray}

\noindent
but this correction is clearly finite, and so must be $\delta_c$. Finally, in the process of gauge fixing we eliminated at the classical level a mixture between $A^{\alpha}$ and $\Pi$ that appeared due to spontaneous breakdown of gauge symmetry. However, at the quantum level, this mixing can in principle reappear. That this is not the case can be deduced by calculating the graphs in Fig.~\ref{Fig6} and verifying that their contributions vanish individually. 

Before continuing with more complicated vertex functions, it is interesting to investigate the general structure of the divergences in the model. To establish its renormalizability at one loop, we have to calculate the superficial degree of divergence $\mathcal{D}$ of an arbitrary diagram $\mathcal{F}$. For this, we will denote the number of the different vertices in the theory as $V_i$, according to the following correspondence,
\begin{eqnarray}
&&(D^{\alpha}\Pi A_{\alpha}\Sigma-D^{\alpha}\Sigma A_{\alpha}\Pi)  \, \longrightarrow \, V_1  
\,\,;\,\, \Pi^2A^2 \, \longrightarrow \, V_2 \nonumber\\
&&\Sigma^2A^2 \, \longrightarrow V_3 \,\,;\,\, 
A^2\Sigma \, \longrightarrow \, V_4 \,\,;\,\, 
\Sigma^4  \, \longrightarrow \, V_5 s\\
&&\Pi^4\, \longrightarrow \, V_6 \,\,;\,\, 
\Sigma^2\Pi^2 \, \longrightarrow \, V_7 \,\,;\,\, 
\Sigma^3 \, \longrightarrow \, V_8 \nonumber\\
&&\Sigma\Pi^2 \, \longrightarrow \, V_9 \,\,;\,\, 
\bar{c}\Sigma c \, \longrightarrow \, V_{10} \nonumber
\end{eqnarray}

\noindent
Let $P_{\Sigma}$, $ P_{A}$, $P_{\Pi}$ and $P_ {c}$ denote the number of propagators for each field in the model. Then, for an arbitrary diagram, the superficial degree of divergence $\mathcal{D}$ is given by
\begin{eqnarray}\label{eq24}
\mathcal{D}=2L-P_A-P_{\Sigma}-P_{\Pi}-P_{c}+{V_1\over 2}~,
\end{eqnarray} 

\noindent
since each loop $L$ furnishes two powers of momenta (remember that the contraction of the loop to a point in $\theta$-space reduces the power counting by one), and each propagator contributes with $-1$ to $\mathcal{D}$~\cite{Gates:1983nr}. By using the topological relations,
\begin{eqnarray}\label{eq23}
2P_{\Sigma}+E_{\Sigma}&=&V_1+2V_2+V_4+4V_5+2V_7+3V_8+V_9+V_{10}~,\nonumber\\
2P_{\Pi}+E_{\Pi}&=&V_1+2V_3+4V_6+2V_7+2V_9~,\\
2P_{A}+E_{A}&=&V_1+2V_2+2V_3+2V_4~,\hspace{1cm}2P_{c}+E_{c}=2V_{10}~,\nonumber
\end{eqnarray}

\noindent
as well as the Euler identity $L+V-P=1$, we find
\begin{eqnarray}\label{eq25}
\mathcal{D}=2-{1\over 2}(E_A+E_{\Sigma}+E_{\Pi}+E_{c})-{1\over 2}(V_4+V_8+V_9+V_{10})-{N_D\over 2}~,
\end{eqnarray}

\noindent
where $N_D$ is the number of covariant derivatives $D^{\alpha}$ acting on the external fields of the diagram.

\begin{figure}[ht]
\includegraphics[]{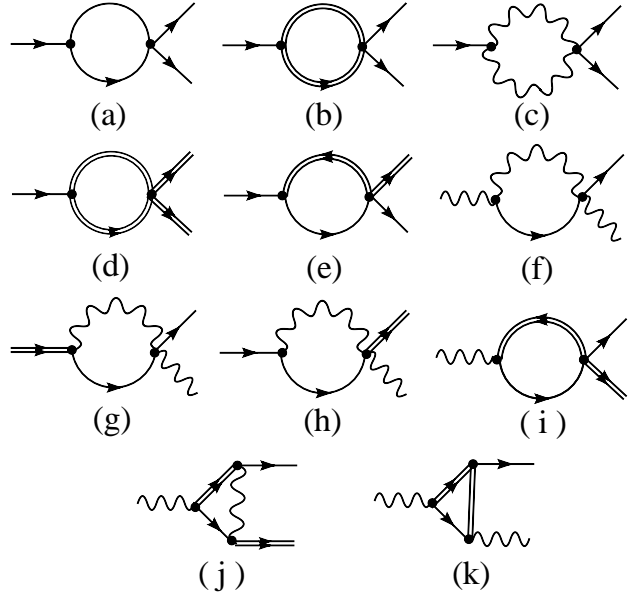}
\caption{Diagrams of three points with logarithmic superficial degree of divergence.} \label{Fig7}
\end{figure}

\begin{figure}[ht]
\includegraphics[]{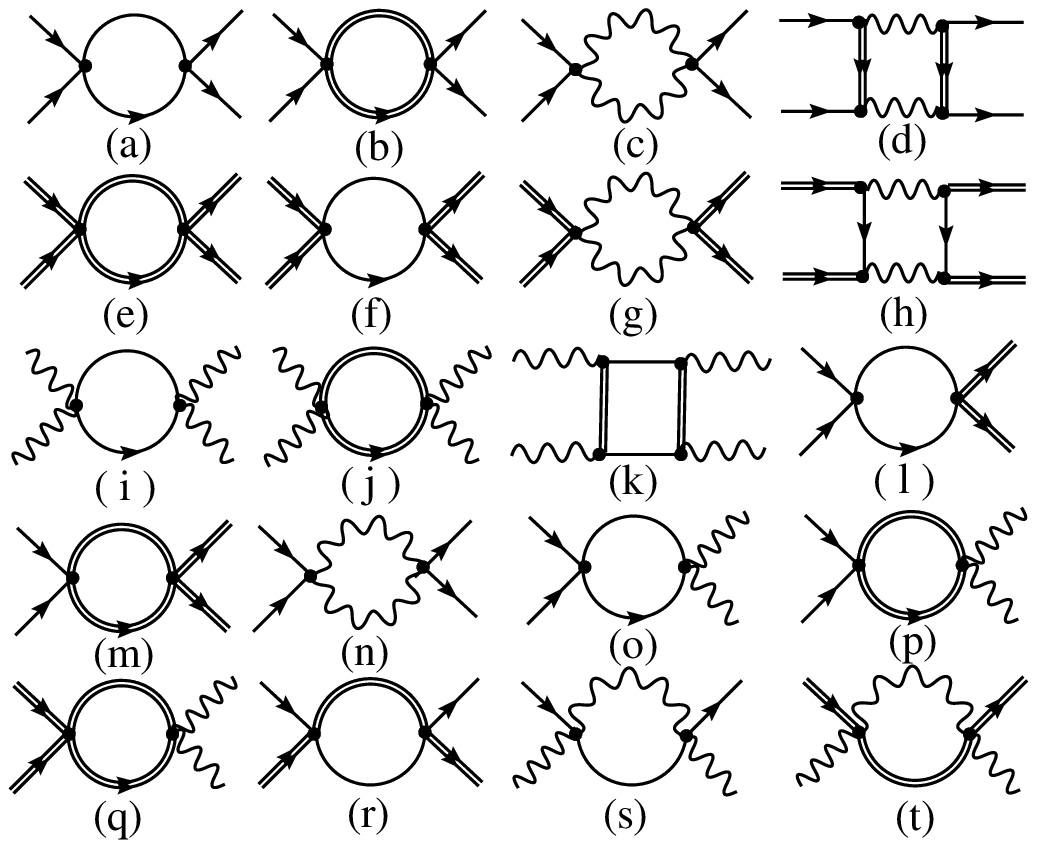}
\caption{Four point diagrams with logarithmic superficial degree of divergence.}  \label{Fig8}
\end{figure}

We can now conclude that, at one loop, any diagram with more than four external legs is finite. No diagram possesses linear divergence except the ones with one or two external legs, which have already been taken into account. Some diagrams with three and four external legs are superficially  logarithmically divergent, and are shown in Figs.~\ref{Fig7} and~\ref{Fig8}.

The diagrams \ref{Fig7}(a-i) possess an expression proportional to
\begin{eqnarray}\label{eq27}
\int\!{d^3k\over(2\pi)^3}d^2\theta_1d^2\theta_2{(D^2+M_1)\delta_{12}(D^2+M_2)\delta_{12}\over[(k+p)^2+M_1^2](k^2+M_2^2)}
\mathcal{G}(p,\theta_1,\theta_2)~,
\end{eqnarray}

\noindent
where $\mathcal{G}(p,\theta_1,\theta_2)$ is the factor involving the external fields. Inspecting the consequences of the D-algebra manipulations on Eq.~(\ref{eq27}), we realise that the only nonvanishing contributions are those where one of the operators $D^2$ is moved to the external fields, and also terms proportional to $M_1$ or $M_2$. Thus, by simple power counting, we can state that those diagrams are finite. Diagram~\ref{Fig7}(j) possess three internal propagators and have the following structure,
\begin{eqnarray}\label{eq28}
\int\!{d^3k\over(2\pi)^3}d^2\theta_1d^2\theta_2d^2\theta_3~(|D|^3_{\rm vertex})\frac{(D^2+M_1)\delta_{12}
(D^2+M_2)\delta_{23}(D^2+M_3)\delta_{31}}{k^6}\,{\mathcal{G}(p,\theta_1,\theta_2,\theta_3)}~,
\end{eqnarray}

\noindent
where $\mathcal{G}(p,\theta_1,\theta_2,\theta_3)$ has the same meaning as above, and the $(|D|^3_{\rm vertex})$ factor represents schematically three supercovariant derivatives arising from the vertices. After D-algebra manipulations, expression~(\ref{eq28}) reduces to something proportional to
\begin{eqnarray}\label{eq29}
\int\!{d^3k\over (2\pi)^3} d^2\theta~ \frac{k_{\alpha\beta}}{k^4}|D| {\mathcal{G}^{\alpha\beta}(p,\theta_1,\theta_2,\theta_3)}~,
\end{eqnarray}

\noindent
which vanishes due to the symmetrical integration in $k$. As for graph~\ref{Fig7}(k), its contribution has a similar structure, lacking one $D$ in comparing with Eq.~(\ref{eq29}). Therefore, the three point functions of the model are finite.  

As for the four-point vertex functions, depicted in Fig.~\ref{Fig8}, they also have superficial degree of divergence $\mathcal{D}=0$, but also turn out to be finite. Indeed, diagrams with two internal propagators possess the same 
general structure of the graphs \ref{Fig7}(a-i) discussed above. As for the graphs with four internal propagators, their divergent parts are proportional to
\begin{eqnarray}\label{eq30}
\int\!{d^3k\over(2\pi)^3} d^2\theta_1d^2\theta_2d^2\theta_3d^2\theta_4~(|D|^4_{vertex})
\frac{D^2\delta_{12}D^2\delta_{23}D^2\delta_{34}D^2\delta_{41}}{k^8}
{\mathcal{G}(p,\theta_1,\theta_2,\theta_3,\theta_4)}~.
\end{eqnarray}

\noindent
The finiteness of these contributions follows from arguments similar to the ones we discussed before.

In summary, in this work we investigated some perturbative aspects of the supersymmetric Chern-Simons field in the presence of spontaneous breaking of $U(1)$ gauge symmetry. The model possesses a classical non-trivial vacuum for $\langle\Phi\rangle = \sqrt{\mu/2\lambda}$, which represents a supersymmetric phase were the gauge symmetry is spontaneously broken. In the calculation of the quantum corrections to the effective action of the model, we have used a $R_{\xi}$ gauge to eliminate the mixture among the $A^{\alpha}$ and $\Pi$ superfields. The absence of this mixing is preserved at one loop level. The renormalization of the gap equations, including one loop corrections, ensures also finiteness of the two-point functions of the $\Sigma$ and $\Pi$ superfields, while the corrections to the quadratic effective action of the gauge superfield $A^{\alpha}$ turns out to be finite. Thus, the only infinite renormalization needed by this model is in the mass counterterm $\delta_{\mu}$. By means of general arguments, we could show that no divergences arise in the vertex functions of up to four points, which are the only having UV divergence at the one loop level, according to the power counting. This completes the proof of the one loop renormalizability.

\vspace{1cm}

\textbf{Acknowledgments}

This work was partially supported by the Brazilian agencies Funda\c{c}\~{a}o de Amparo 
\`{a} Pesquisa do Estado de S\~{a}o Paulo (FAPESP) and Conselho 
Nacional de Desenvolvimento Cient\'{\i}fico e Tecnol\'{o}gico (CNPq).

\begin{thebibliography}{99}
\bibitem{footnote}
In the more complicated cases, the D-algebra manipulations were performed using a computer algebra software, see~\cite{susymath}.
\end{thebibliography}
\end{document}